\documentstyle[aps,preprint,pra,url]{revtex}
\begin{document}
\newcommand{\etal}{{\em et al.}\/}
\newcommand{\IP}{inner polarization}
\newcommand{\IPF}{\IP\ function}
\newcommand{\IPFs}{\IP\ functions}
\newcommand{\ket}[1]{$\left|#1\right>$}
\newcommand{\auth}[2]{#1 #2, }
\newcommand{\jcite}[4]{#1 {\bf #2}, #3 (#4)}
\newcommand{\et}{ and }
\newcommand{\twoauth}[4]{#1 #2 and #3 #4,}
\newcommand{\oneauth}[2]{#1 #2,}
\newcommand{\andauth}[2]{and #1 #2, }
\newcommand{\book}[4]{{\it #1} (#2, #3, #4)}
\newcommand{\erratum}[3]{\jcite{erratum}{#1}{#2}{#3}}
\newcommand{\inpress}[1]{{\it #1}}
\newcommand{\inbook}[5]{In {\it #1}; #2; #3: #4, #5}
\newcommand{\JCP}[3]{\jcite{J. Chem. Phys.}{#1}{#2}{#3}}
\newcommand{\jms}[3]{\jcite{J. Mol. Spectrosc.}{#1}{#2}{#3}}
\newcommand{\jmsp}[3]{\jcite{J. Mol. Spectrosc.}{#1}{#2}{#3}}
\newcommand{\jmstr}[3]{\jcite{J. Mol. Struct.}{#1}{#2}{#3}}
\newcommand{\cpl}[3]{\jcite{Chem. Phys. Lett.}{#1}{#2}{#3}}
\newcommand{\cp}[3]{\jcite{Chem. Phys.}{#1}{#2}{#3}}
\newcommand{\pr}[3]{\jcite{Phys. Rev.}{#1}{#2}{#3}}
\newcommand{\jpc}[3]{\jcite{J. Phys. Chem.}{#1}{#2}{#3}}
\newcommand{\jpcA}[3]{\jcite{J. Phys. Chem. A}{#1}{#2}{#3}}
\newcommand{\jpca}[3]{\jcite{J. Phys. Chem. A}{#1}{#2}{#3}}
\newcommand{\jpcB}[3]{\jcite{J. Phys. Chem. B}{#1}{#2}{#3}}
\newcommand{\jpB}[3]{\jcite{J. Phys. B}{#1}{#2}{#3}}
\newcommand{\PRA}[3]{\jcite{Phys. Rev. A}{#1}{#2}{#3}}
\newcommand{\PRB}[3]{\jcite{Phys. Rev. B}{#1}{#2}{#3}}
\newcommand{\PRL}[3]{\jcite{Phys. Rev. Lett.}{#1}{#2}{#3}}
\newcommand{\jcc}[3]{\jcite{J. Comput. Chem.}{#1}{#2}{#3}}
\newcommand{\molphys}[3]{\jcite{Mol. Phys.}{#1}{#2}{#3}}
\newcommand{\mph}[3]{\jcite{Mol. Phys.}{#1}{#2}{#3}}
\newcommand{\APJ}[3]{\jcite{Astrophys. J.}{#1}{#2}{#3}}
\newcommand{\cpc}[3]{\jcite{Comput. Phys. Commun.}{#1}{#2}{#3}}
\newcommand{\jcsfii}[3]{\jcite{J. Chem. Soc. Faraday Trans. II}{#1}{#2}{#3}}
\newcommand{\FD}[3]{\jcite{Faraday Discuss.}{#1}{#2}{#3}}
\newcommand{\prsa}[3]{\jcite{Proc. Royal Soc. A}{#1}{#2}{#3}}
\newcommand{\jacs}[3]{\jcite{J. Am. Chem. Soc.}{#1}{#2}{#3}}
\newcommand{\joptsa}[3]{\jcite{J. Opt. Soc. Am.}{#1}{#2}{#3}}
\newcommand{\cjc}[3]{\jcite{Can. J. Chem.}{#1}{#2}{#3}}
\newcommand{\ijqcs}[3]{\jcite{Int. J. Quantum Chem. Symp.}{#1}{#2}{#3}}
\newcommand{\ijqc}[3]{\jcite{Int. J. Quantum Chem.}{#1}{#2}{#3}}
\newcommand{\spa}[3]{\jcite{Spectrochim. Acta A}{#1}{#2}{#3}}
\newcommand{\tca}[3]{\jcite{Theor. Chem. Acc.}{#1}{#2}{#3}}
\newcommand{\tcaold}[3]{\jcite{Theor. Chim. Acta}{#1}{#2}{#3}}
\newcommand{\jpcrd}[3]{\jcite{J. Phys. Chem. Ref. Data}{#1}{#2}{#3}}
\newcommand{\science}[3]{\jcite{Science}{#1}{#2}{#3}}
\newcommand{\CR}[3]{\jcite{Chem. Rev.}{#1}{#2}{#3}}
\newcommand{\bbpc}[3]{\jcite{Ber. Bunsenges. Phys. Chem.}{#1}{#2}{#3}}
\newcommand{\acie}[3]{\jcite{Angew. Chem. Int. Ed.}{#1}{#2}{#3}}
\newcommand{\ijck}[3]{\jcite{Int. J. Chem. Kinet.}{#1}{#2}{#3}}
\newcommand{\jct}[3]{\jcite{J. Chem. Thermodyn.}{#1}{#2}{#3}}

\newcommand{\deltah}[0]{$\Delta$H}
\newcommand{\deltahf}[0]{$\Delta$H$_f$}
\newcommand{\hoof}[0]{$\Delta$H$_f$$^{298}$}
\newcommand{\abin}{{\em ab initio}}
\newcommand{\hof}{$\Delta H^\circ_f$}
\newcommand{\hofzero}{$\Delta H^\circ_{f,0}$}
\newcommand{\HOF}{$\Delta H^\circ_{f,298}$}

\draft
\title{Definitive heat of formation of methylenimine, CH$_2$=NH, and of 
methylenimmonium ion, CH$_2$NH$_2^+$, by means of W2 theory\thanks{Dedicated to Prof. Paul von Ragu\'e Schleyer on the occasion of his 70th birthday}}
\author{Gl\^enisson de Oliveira\thanks{Present address: 
Chemistry Department, Pensacola Christian College, 250 Brent Lane, 
Pensacola, FL 32503}, Jan M.L. Martin\thanks{Author to whom correspondence should
be addressed. {\rm E-mail:} {\tt comartin@wicc.weizmann.ac.il}}, and Indira K.C. Silwal}
\address{Department of Organic Chemistry,
Weizmann Institute of Science,
IL-76100 Re\d{h}ovot, Israel
}
\author{Joel F. Liebman}
\address{Department of Chemistry and Biochemistry,
University of Maryland, Baltimore County (UMBC), Baltimore, Maryland 21250}
\date{Submitted to {\it J. Comput. Chem.} Dec. 14, 2000; Accepted Dec. 17, 2000}
\maketitle
\begin{abstract}
A long-standing controversy concerning the heat of formation of methylenimine
has been addressed by means of the W2 (Weizmann-2) thermochemical approach.
Our best calculated values, \HOF(CH$_2$NH)=21.1$\pm$0.5
kcal/mol and \HOF(CH$_2$NH$_2^+$)=179.4$\pm$0.5 kcal/mol, are in good agreement with the
most recent measurements but carry a much smaller uncertainty. As a by-product, we
obtain the first-ever accurate anharmonic force field for methylenimine: upon
consideration of the appropriate resonances, the experimental gas-phase band origins
are all reproduced to better than 10 cm$^{-1}$. Consideration
of the difference between a fully anharmonic zero-point vibrational energy and 
B3LYP/cc-pVTZ harmonic frequencies scaled by 0.985 suggests that the calculation
of anharmonic zero-point vibrational energies can generally be dispensed with,
even in benchmark work, for rigid molecules.
\end{abstract}
\newpage
\section{Introduction}
Methylene imine (methanimine, formaldimine, CH$_2$=NH) is a pyrolysis product
of amines\cite{Joh72} as well as a photolysis product of methylazide\cite{Mil61}
and diazomethane\cite{Moo65}.
It has obvious chemical importance as the simplest imine\cite{patai}, 
and with its carbon-nitrogen double bond, methylenimine is a bonding 
paradigm for numerous nitrogen-containing heterocycles.
The molecule is 
also of astrophysical interest, 
having been detected in dark interstellar dust clouds\cite{Godf73}.

The heat of formation of methylenimine (methanimine, formaldimine, CH$_2$=NH), 
is the subject of a
long-standing controversy in the literature.
Experimental values for the heat of formation (\hof) of this molecule span a 
range of 10 kcal/mol, and have fairly large experimental uncertainties (about 2 to 3 kcal/mol). 
The first experiment, by DeFrees and Hehre in 1978,\cite{DeF78}
used the bracketing method to evaluate the hydride affinity of HCNH$^+$, 
and hence derived \HOF[CH$_2$=NH]=26.4$\pm$3.2 kcal/mol. Ten years later,
Grela and Colussi\cite{Gre88} obtained a value of 25$\pm$3 kcal/mol from
the deprotonation reaction of CH$_3$NH. A Moscow group had meanwhile 
obtained\cite{russians} 21$\pm$4 kcal/mol by photoionization mass spectrometry 
of pyrolysis products of azetidine. In 1990, Peerboom, Ingemann, Nibbering,
and Liebman (PINL) \cite{Pee90}
bracketed the proton affinity (PA) of CH$_2$NH as 204$\pm$2 kcal/mol by means
of ion cyclotron resonance: in combination with an earlier determination
of the heat of formation of CH$_2$NH$_2^+$ (177--178 kcal/mol) by Lossing et al.\cite{Los81}
from appearance energy measurements, they obtained
\HOF[CH$_2$=NH]=16.5$\pm$2 kcal/mol. Then in 1992, 
Holmes et al.\cite{Hol92} determined the ionization potential of the CH$_2$NH radical by means of
energy-resolved electron impact, and derived an upper limit of
22$\pm$3 kcal/mol for \HOF[CH$_2$=NH]: they propose an `evaluated'
\HOF[CH$_2$=NH]=21$\pm$4 kcal/mol, which happens to be identical
to the Moscow group value. Around the same time, Smith, Pople,
Curtiss, and Radom (SPCR)\cite{Smi92} carried out a computational study
in which reaction energies for ten reactions involving CH$_2$NH were
computed by means of G2 theory\cite{g2}: in combination with experimental
data for the auxiliary species\cite{liebnote}, they obtained \HOF=20.6$\pm$2.4 kcal/mol
averaged over the ten reactions. Although this is in good agreement with the 
Holmes et al. experiment, the error bars are still a far cry from `chemical 
accuracy' (1 kcal/mol).

Very recently, two of us proposed\cite{w1} two new computational 
thermochemistry methods known as W1 and W2 (Weizmann-1 and Weizmann-2) theory which,
for molecules dominated by dynamical correlation, yield heats
of formation to within 0.25 kcal/mol (1 kJ/mol) accuracy, on average.
A subsequent validation
study\cite{w1w2validate} for a much larger data set came to similar 
conclusions. Since CH$_2$NH is still small enough for a W2 calculation
to be carried out on fast workstation computers, this would appear to
be the tool of choice for resolving the controversy on its heat of formation
for once and for all. This is the primary purpose of the present paper.

As a by-product, we shall obtain an accurate ab initio anharmonic force 
field for CH$_2$NH. (For the highest possible accuracy, it is {\em in principle} advisable
to obtain the molecular zero-point vibrational energy ZPVE from an anharmonic force field
rather than from scaled harmonic frequencies.)
Aside from matrix isolation work\cite{Mil61,Jac75}, a 
respectable amount of high-resolution IR data is available for this molecule.
Following early high-resolution work by Johnson and
Lovas\cite{JoLo}, Allegrini et al.\cite{All79} obtained a high-resolution $\nu_4$
by CO laser Stark spectroscopy. Duxbury and Le Lerre\cite{Dux82} studied the
$\nu_5$ and $\nu_6$ bands by Fourier transform IR (FTIR) spectroscopy, including
an analysis of the Coriolis interaction (along the $c$ axis) between those modes.
The $\nu_7$, $\nu_8$, and $\nu_9$ modes, as well as the strong Coriolis
interactions between them, were studied by Halonen and Duxbury\cite{Hal85a}, while
these same authors studied $\nu_2$ and the ($\nu_3$,$2\nu_5$) Fermi resonant
band pair in a companion paper\cite{Hal85b} and reported $\nu_1$ elsewhere\cite{Hal85c}.

General harmonic force fields were derived by Jacox and Milligan\cite{Jac75}, 
by Hamada et al.\cite{Ham84} and by
Halonen, Deeley, and Mills\cite{Hal86}: the latter authors also remeasured and reanalyzed the 
$(\nu_7,\nu_8,\nu_9)$ triad. (A microwave substitution structure was obtained
by Pearson and Lovas\cite{Pea77}.) 
To the authors' knowledge, the only anharmonic force field available
is a comparatively low-level (MP2/6-311G**) ab initio calculation by Pouchan and Zaki\cite{Pou97}.
Extensive experience has shown (e.g. \cite{c2h4} and references therein) that the CCSD(T) 
(coupled cluster with all single and double substitutions\cite{Pur82} and a quasiperturbative
correction for connected triple excitations\cite{Rag89}) electron correlation method in
conjunction with a basis set of $spdf$ quality generally yields computed fundamentals
within better than 10 cm$^{-1}$ of the observed gas-phase values. Thus, obtaining a force
field of such quality is the secondary purpose of the present note.

\section{Computational methods}

Geometry optimizations and vibrational frequency
calculations using the B3LYP (Becke 3-parameter-Lee-Yang-Parr\cite{Bec93,LYP})
density functional method have been carried out using Gaussian 98 revision A7\cite{g98revA7}.  All
other calculations were carried out using MOLPRO 98.1\cite{molpro98}, and a driver for the W1/W2 calculations\cite{autoW1W2}
written in MOLPRO's scripting language,  running on Compaq XP1000
and Compaq ES40 computers in our research group.

W1 and W2 theory are described in detail elsewhere\cite{w1,w1w2validate}.
Briefly, both methods involve separate extrapolations to the infinite-basis
limit, using a sequence of Dunning correlation consistent\cite{Dun89,Dun97}
(cc-pV$n$Z)
and augmented correlation consistent\cite{Ken92} (aug-cc-pV$n$Z) 
basis sets, of the SCF, CCSD valence correlation, and (T) valence correlation
energies. In addition, contributions of inner-shell correlation and 
(Darwin and mass-velocity\cite{Cow76,Mar83}) 
scalar relativistic effects are obtained at the CCSD(T) and ACPF
(averaged coupled pair functional\cite{Gda88}) levels with the MTsmall
basis set\cite{w1}. While the more economical W1 theory uses a
B3LYP/cc-pVTZ reference geometry and extrapolates from aug$'$-cc-pV$n$Z
(n=D,T,Q) basis sets, the more expensive (and rigorous) W2 theory
employs a CCSD(T)/cc-pVQZ reference geometry and aug$'$-cc-pV$n$Z
($n$=T,Q,5) basis sets. (Regular cc-pV$n$Z
basis sets are used throughout on hydrogen.) In addition, we considered
W1h and W2h results, where the `h' (for 'hetero-atom') indicates that
augmented basis sets are only used on group V, VI, and VII elements and
not on group III and IV elements. 

The largest basis set CCSD calculations in W2 and W2h theory were carried
out using the direct CCSD implementation\cite{dirccsd} of Lindh, Sch\"utz,
and Werner as present in MOLPRO 98.1. All energies for the open-shell 
separated atoms were obtained using the restricted open-shell CCSD(T)
energy as defined in Ref.\cite{Wat93}.

For comparison, we shall also present data for the isoelectronic
C$_2$H$_4$ and N$_2$H$_2$ molecules. 

A complete CCSD(T)/cc-pVTZ quartic force field for CH$_2$NH was generated
in internal coordinates (four stretches, three bends, two torsions). 
Internal coordinate geometries were generated by recursive application
of the central difference formula to the coordinates being differentiated,
with step sizes of 0.01 \AA\ or radian around the minimum energy geometry 
being used. Cartesian coordinates for this `grande list' of points 
were generated using INTDER\cite{intder}: this list of geometries was reduced to a `petite list'
of unique points by means of comparison of sorted distance matrices. Thus,
2241 points in $C_s$ symmetry and 460 additional points in $C_1$ symmetry are
obtained. Since this type of application is a textbook example of an
`embarrassingly parallel'\cite{mevich} computational problem, the energy calculations were
carried out on a 26-node experimental PC-farm at the Department of 
Particle Physics,
Weizmann Institute of Science. In order to minimize roundoff error in
the finite differentiation, integral evaluation cutoffs as well as
SCF and CCSD convergence criteria were tightened such that the energies
are obtained to essentially machine precision. Quartic contamination
was removed from the quadratic force constants. The final internal 
coordinate force field was transformed to Cartesian coordinates using
INTDER, and transformed to normal coordinates as well as subjected to
a standard second-order rovibrational perturbation theory (VIB-PT2) analysis\cite{Pap82}
using SPECTRO\cite{spectro} and POLYAD\cite{polyad}.

\section{Results and Discussion}

\subsection{Anharmonic force field}

A plethora of resonances exists involving the three XH stretching modes ${\nu_1,\nu_2,\nu_3}$
on the one hand and two-quantum states within the ${\nu_4,\nu_5,\nu_6}$ block on
the other hand (the modes involved are the CN stretch, the HNC bend, and the HCH scissoring
mode, respectively). For this reason, we deperturbed the anharmonic constants
for all resonances of the type $\nu_x\approx\nu_y+\nu_z$ (where x=\{1,2,3\} and
y,z=\{4,5,6\}, e.g. $\nu_2\approx\nu_4+\nu_6$ or $\nu_3\approx2\nu_5$), and set up and diagonalized a $9\times9$ resonance matrix
involving all these states. (Formulas for the various higher-order resonance
matrix elements were taken from Ref.\cite{Mar97}.) The resonance matrix and its eigensolution
are given in Table \ref{tab:freq}, while the computed harmonic frequencies and fundamentals (as well
as any bands in resonance with them) are given in Table \ref{tab:eigen}, compared with experiment
and with results from the previous lower-level
(MP2/6-311G**) calculation by Pouchan and Zaki\cite{Pou97}. A complete force field
and sets of spectroscopic constants are available as supplementary material to the
present paper.

First of all, as readily seen from the solution of the $9\times9$ resonance matrix,
the $\nu_3\approx2\nu_5$ resonance is so severe that the two perturbed states
are basically 50:50 mixtures of the respective deperturbed states, and that
an assignment of an observed band to either $\nu_3$ or $2\nu_5$ is somewhat
academic. Similar remarks apply to the $\nu_2\approx\nu_4+\nu_6$ resonance: in both cases,
the assignments in the table were made based on the ordering of the deperturbed states.
The $\nu_1\approx2\nu_4$
resonance is also quite severe but an unambiguous assignment is still possible there. 
(For a system like this, a full nine-dimensional solution by variational methods\cite{Bowman}
or high-order canonical Van Vleck perturbation theory\cite{cvpt} would certainly
be helpful: this is however
beyond the scope of the present study since we are primarily interested in the
thermochemistry.)

This being said, agreement between computed and observed vibrational band origins
is basically as good as we can reasonably expect at this level of theory, with
all computed-observed discrepancies lying below 10 cm$^{-1}$.  The very good 
agreement between the present band origins and the earlier lower-level results
is somewhat fortuitous, given the discrepancies of up to 50 cm$^{-1}$ between
the two sets of {\em harmonic} frequencies. It has been our experience that
MP2 computed anharmonicities for XH stretching modes tend to be seriously
overestimated, and the present system forms no exception.

The only experimental equilibrium geometry available from the literature
is a microwave ($r_s$) substitution structure\cite{Pea77}. Agreement between our
calculations and the $r_s$ geometry is as good as we can reasonably expect (Table \ref{tab:geom}).
The effect of correlating the (1s)-like inner-shell electrons on the geometry
follows expected trends\cite{cc} (Table \ref{tab:geom}). (The MTsmall core correlation basis set\cite{w1}
as used in W1 and W2 theory was employed for this purpose.) No experimental $r_e$ or 
$r_z$ geometry is available, but an indirect measure of the quality of our computed
CCSD(T)/MTsmall geometry can be obtained by substituting it in the VIB-PT2 analysis
and comparing the ground-state rotational constants thus obtained with their precisely
known\cite{Hal85a} experimental counterparts. Our computed $A_0$=6.54242, $B_0$=1.15615, $C_0$=0.97936
cm$^{-1}$ agree to better than 0.1\% with the observed values\cite{Hal85a} 6.544896(1), 1.1555459(1), and
0.9790851(1) cm$^{-1}$: given the quadratic dependence of the rotational constants on the geometry,
this in fact suggests an even better agreement between the underlying computed $r_e$ geometry and Nature.
In order to assist future experimental work on the protonated species CH$_2$NH$_2^+$, we have 
computed its geometry at the same level (Table \ref{tab:geom}).

For the mode pairs in Coriolis resonance, the computed interaction constants $\xi_{79}^a$=4.701, 
$\xi_{79}^b$=-0.315, $\xi_{78}^a$=3.630, and $\xi_{78}^b$=1.918 cm$^{-1}$ are in fair
agreement with the experimental values\cite{Hal85a} 4.529(1), -0.3305(1), 4.212(1), and 1.8125(1)
cm$^{-1}$, respectively. The computed $\xi_{56}^c$=0.552 cm$^{-1}$ is likewise in reasonable
agreement with the observed value\cite{Dux82} of 0.6911(1) cm$^{-1}$.

Let us finally turn to the zero-point vibrational energy (ZPVE). Our computed value from the CCSD(T)/cc-pVTZ
quartic force field, and including the $E_0$ correction, is 24.69 kcal/mol. As seen
in Table \ref{tab:heat}, this differs by no more than 0.10 kcal/mol from the zero-point correction
used in W1 theory, i.e. B3LYP/cc-pVTZ harmonic frequencies scaled by 0.985. The same
remark holds true (Table \ref{tab:heat}) for the isoelectronic molecules C$_2$H$_4$ (difference +0.04
kcal/mol) and trans-HNNH (difference +0.03 kcal/mol), for which large basis set
CCSD(T) quartic force fields are available from previous work\cite{c2h4more,n2h2}. 
There are certainly situations (e.g. nonrigid molecules --- i.e. those exhibiting
low-barrier internal rotations and/or inversions, or very low frequency modes ---
or very anharmonic systems such as $H_3^+$)
where anything less than an anharmonic force field
is fundamentally inappropriate for the zero-point energy. Yet it would appear to
be that the immensely less expensive scaled harmonic B3LYP ZPVE is appropriate even
for benchmark work: any situation where accuracy of $\pm$0.1 kcal/mol on the 
computed atomization energy is essential is presently beyond direct treatment even
by W2 theory.

\subsection{Heat of formation}

Total atomization energies (TAE$_e$ if zero-point exclusive, TAE$_0$ at 0 K)
at the W1, W1h, W2, and W2h levels for
CH$_2$NH, CH$_2$NH$_2^+$, trans-HNNH, and C$_2$H$_4$ are given 
in Table \ref{tab:heat}, together with a breakdown by components of the
results at the highest level of theory (W2).

In the light of our observation above,
and in order to achieve consistency 
among the species considered (including CH$_2$NH$_2^+$, for which no
anharmonic force field is available), all heats of formation reported in Table
\ref{tab:heat} use the scaled B3LYP ZPVEs rather than their anharmonic counterparts.

At the highest level of theory, we obtain \HOF(CH$_2$NH)=21.1 kcal/mol.
The W2h result is essentially identical; the W1 and W1h results are
slightly different, but still by less than 0.2 kcal/mol.
In the original W2 paper\cite{w1}, the mean absolute error for a sample
of some 30 very accurately known total atomization energies 
was 0.23 kcal/mol; we shall conservatively take our error bar to be
twice that amount, or (after roundoff) $\pm$0.5 kcal/mol.

Our calculation stays below the Holmes et al. upper limit and is in
excellent agreement with both the Holmes et al. evaluated \HOF and
the G2-thermodynamic cycle derived value of SPCR: of course, our
error bar is an order of magnitude smaller than the former and several
times smaller than the latter. 

What is the source of the 5 kcal/mol disagreement between these values and the
earlier PINL measurement? In order to shed light upon this question, we
calculated the heat of formation of protonated methylenimine, and hence
also the PA of the latter compound. At the W1 and W2 levels, we find PA(CH$_2$NH)=207.8
and 207.5 kcal/mol, respectively: this is a minor exception to the rule\cite{w1w2validate}
that W1 and W2 theory
yield essentially identical proton affinities. Our W2 PA is 3.5 kcal/mol
higher than the bracketed value of 204$\pm$2 kcal/mol, but lies within the error bar of the
very recent Bouchoux and Salpin\cite{Bou99} value,
206.2$\pm$1.5 kcal/mol, obtained by the thermokinetic method.\cite{Bou96}
We note that the
accuracy of the bracketing is, by its very nature, in turn dependent on
the accuracy of the PAs of the bases involved in the bracketing experiments.
For the five bases used, namely pyrrole, diisopropyl ether, ammonia,
styrene, and diethyl ether, the PA values from the 1988 Lias et al.
compilation\cite{Lia88} employed by PINL 
differ by up to 2.5 kcal/mol from the more recent
1998 compilation of Hunter and Lias\cite{Hun98}. 

As for the heat of formation of CH$_2$NH$_2^+$, our computed W1 and W2 values are
1.0 and 1.4 kcal/mol higher, respectively, than the value of Lossing et al.\cite{Los81} used
by PINL. Hammerum and S{\o}lling (HS)\cite{Ham99} recently re-evaluated
the experimental data of Lossing et al., using the
298 K enthalpy contributions of Traeger and McLoughlin\cite{Tra81} to
convert the reported threshold energy measurements into heats of formation
at 298 K. In this manner, the value found for methylenimine is
179.7 kcal/mol, in excellent agreement with our W2 calculated
result of 179.4$\pm$0.5 kcal/mol.
HS also calculated the heat of formation of CH$_2$NH$_2^+$
at the G2(MP2) and CBS-Q levels, and found 179.0 and 180.2 kcal/mol,
respectively. For CH$_2$NH, the corresponding values are
20.8 and 22.0 kcal/mol; agreement with our W1 and W2 values is as good
as can reasonably be expected for the G2(MP2) and CBS-Q methods.

W2 is most reliable for molecules that are dominated by dynamical 
correlation energy. One index which we found to be very reliable
for this purpose is the percentage of the binding energy that is
recovered at the SCF level. For CH$_2$NH this is found to be 70\%,
which is closer to C$_2$H$_4$ (77\%) and to molecules essentially
devoid of nondynamical correlation at their equilibrium geometry 
(e.g. water, H$_2$) than to trans-HNNH (52\%, comparable to N$_2$)
which is in a regime of moderate nondynamical correlation. For CH$_2$NH$_2^+$,
SCF accounts for about 63\% of the binding energy.

Interestingly, the reaction
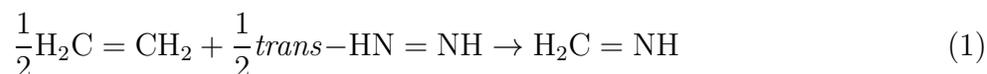
\begin{equation}
\frac{1}{2} {\rm H}_2{\rm C=CH}_2 + \frac{1}{2} {\it trans-}{\rm HN=NH} \rightarrow {\rm H}_2{\rm C=NH}\nonumber
\end{equation}
is computed to be exergonic by about 9 kcal/mol.
At the SCF level, this is even slightly larger: electron 
correlation reduces the difference by about 1.1 kcal/mol.
For comparison, the reaction 
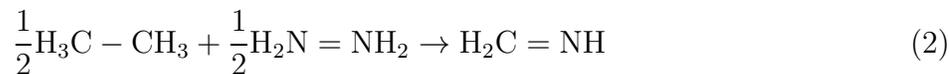
\begin{equation}
\frac{1}{2} {\rm H}_3{\rm C-CH}_3 + \frac{1}{2} {\rm H}_2{\rm N=NH}_2 \rightarrow {\rm H}_2{\rm C=NH}\nonumber
\end{equation}
is exergonic by 6.7 kcal/mol\cite{trcfootnote}, while
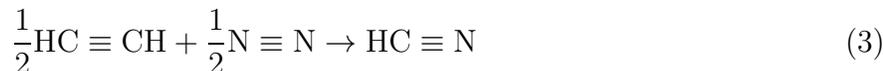
\begin{equation}
\frac{1}{2} {\rm HC\equiv CH} + \frac{1}{2}{\rm N\equiv N} \rightarrow {\rm HC\equiv N}\nonumber
\end{equation}
is {\em endergonic} by 6.9 kcal/mol\cite{hcn}.

\section{Conclusions}

We have calculated benchmark heats of formation of methylenimine and protonated
methylenimine by means of W2 theory. Our best calculated values, \HOF(CH$_2$NH)=21.1$\pm$0.5
kcal/mol and \HOF(CH$_2$NH$_2^+$)=179.4$\pm$0.5 kcal/mol, are in good agreement with the
most recent measurements but carry a much smaller uncertainty. As such, they resolve
a long-standing experimental controversy.

As for many other systems, the difference between a fully anharmonic ZPVE (zero-point
vibrational energy) and a simple B3LYP/cc-pVTZ calculation scaled by 0.985 is negligible
for our purposes.

A first-ever high-quality quartic force field for CH$_2$NH has been made available.
Agreement with experimental high-resolution IR data is better than 10 cm$^{-1}$.
Reliable band origins for the stretching modes require diagonalization of a 
$9\times9$ resonance polyad involving $\nu_x+\nu_y$ ($x,y$=4,5,6).

\acknowledgments
GdO was a Postdoctoral Fellow of the Feinberg 
Graduate School (Weizmann Institute) when this work was initiated. 
JM is the incumbent of the Helen 
and Milton A. Kimmelman Career Development Chair. 
The Weizmann Institute Pilot PC Farm was used courtesy of 
the Department of Particle Physics and the Weizmann Institute 
Computing Center.
JL thanks the US National
Institute of Standards and Technology for partial support of his
thermochemical studies.

\subsection*{Supplementary material} 
The CCSD(T)/cc-pVTZ quartic force field of methylenimine
in internal and in normal coordinates is available on the
World Wide Web at the Uniform Resource Locator
(URL) \url{http://theochem.weizmann.ac.il/web/papers/ch2nh.html}

\begin{table}
\caption{Computed harmonic frequencies $\omega_i$ of CH$_2$NH; computed and
observed fundamental frequencies and selected overtones and combination
bands of CH$_2$NH. All quantities in cm$^{-1}$.\label{tab:freq}}
\begin{tabular}{ldddrdd}
$i$ & $\omega_i$ & $\omega_i$ & $\nu_i$ & $\nu_i$ & $\nu_i$ & $\nu_i$ \\
    &  CCSD(T)/  & MP2/       & Expt.     & Expt.     & CCSD(T)/ & MP2/     \\
    &  cc-pVTZ   & 6-311G**   & gas phase & Ar matrix$^a$ & cc-pVTZ  & 6-311G** \\
\hline
1      & 3440.5& 3491.5 & 3262.622$^b$ &      &    3268.5 &  3260.7  \\
2      & 3152.1& 3198.3 & 3024.452$^c$ & 3036 &    3024.7 &  3017.4  \\
3      & 3052.0& 3086.8 & 2914.184$^c$ & 2926 &    2914.4 &  2903.1  \\
4      & 1675.3& 1685.1 & 1638.30$^d$ & 1641 &    1634.8 \\
5      & 1483.2& 1510.1 & 1452.039$^e$ & 1453 &    1451.9 \\
6      & 1386.9& 1397.2 & 1344.267$^e$ & 1348 &    1350.1 \\
7      & 1073.0& 1089.4 & 1058.181$^f$ & 1059 &    1058.1 \\
8      & 1162.0& 1186.4 & 1126.988$^f$ & 1123 &    1131.6 \\
9      & 1080.0& 1101.9 & 1060.760$^f$ & 1063 &    1059.3 \\
$2\nu_4$      &&        &           &     &     3257.1 & 3258.3 \\
$2\nu_5$      &&        &  2884.986$^c$&     &     2878.4 & 2886.7 \\
$2\nu_6$      &&        &           &     &     2685.5 & ---    \\
$\nu_4+\nu_5$ &&        &           &     &     3080.9 & 3064.2 \\
$\nu_4+\nu_6$ &&        &           &     &     2958.3 & 2944.2 \\
$\nu_5+\nu_6$ &&        &           &     &     2796.1 & 2781.4 \\
\end{tabular}

(a) Refs.\cite{Mil61,Jac75}
(b) Ref.\cite{Hal85c}
(c) Ref.\cite{Hal85b}
(d) Ref.\cite{Dux81}
(e) Ref.\cite{Dux82}
(f) Ref.\cite{Hal85a}
\end{table}

\begin{table}
\caption{Resonance matrix involving the stretching modes and eigensolution. Units are cm$^{-1}$
except for the eigenvectors, which are dimensionless.\label{tab:eigen}}
\squeezetable
\begin{tabular}{lddddddddd}
\ket{100000000} & 3255.16 \\
\ket{010000000} &   -2.982  & 2992.19 \\
\ket{001000000} &   -1.978  &    4.261  & 2901.63 \\
\ket{000110000} &   -3.938  &   12.704  &   20.856  & 3076.43 \\
\ket{000101000} &  -17.022  &  -35.988  &   16.062  &   -1.728  & 2981.60 \\
\ket{000011000} &   -8.823  &   30.951  &  -10.313  &   -2.584  &    2.868  & 2802.89 \\
\ket{000200000} &   -5.229  &   -7.531  &  -28.565  &   -1.577  &   -3.948  &   -0.131  & 3256.70 \\
\ket{000020000} &   -0.409  &  -12.821  &  -19.947  &    2.550  &   -0.511  &   -1.473  &   -2.532  & 2901.09 \\
\ket{000002000} &  -75.390  &   -4.853  &  -11.136  &   -0.478  &   -4.373  &    0.066  &   -1.828  &   -0.425  & 2696.37 \\
\hline
Eigenvectors:\\
\hline
                &$2\nu_6$ &$\nu_5+\nu_6$&$2\nu_5$ & $\nu_3$ & $\nu_4+\nu_6$ & $\nu_2$ & $\nu_3+\nu_4$ & $2\nu_4$ & $\nu_1$ \\
                &  2685.5 & 2796.1 &2878.4 &2914.4 &2958.3 &3024.7 &3080.9 &3257.2 &3268.5 \\
\hline
\ket{100000000} &  -0.132 & -0.014 & 0.010 & 0.019 &-0.040 & 0.034 &-0.017 & 0.434 & 0.889 \\
\ket{010000000} &  -0.018 &  0.167 &-0.056 & 0.217 &-0.579 &-0.746 &-0.172 &-0.027 & 0.008 \\
\ket{001000000} &  -0.052 & -0.112 &-0.719 &-0.612 &-0.261 & 0.082 &-0.108 &-0.076 & 0.031 \\
\ket{000110000} &   0.001 & -0.008 & 0.090 & 0.055 & 0.088 & 0.165 &-0.976 &-0.027 &-0.010 \\
\ket{000101000} &  -0.022 &  0.056 & 0.089 & 0.268 &-0.718 & 0.627 & 0.065 &-0.039 &-0.044 \\
\ket{000011000} &  -0.009 & -0.978 & 0.088 & 0.112 &-0.113 &-0.104 &-0.005 &-0.009 &-0.017 \\
\ket{000200000} &  -0.007 & -0.004 &-0.058 &-0.037 &-0.047 &-0.001 &-0.032 & 0.894 &-0.438 \\
\ket{000020000} &  -0.008 & -0.014 &-0.671 & 0.699 & 0.236 & 0.066 & 0.011 &-0.002 & 0.000 \\
\ket{000002000} &  -0.989 &  0.012 & 0.041 & 0.013 & 0.045 &-0.008 & 0.009 &-0.059 &-0.116 \\
\end{tabular}
\end{table}

\begin{table}
\caption{Computed and observed equilibrium geometries for CH$_2$NH and CH$_2$NH$_2^+$\label{tab:geom}}
\squeezetable
\begin{tabular}{lddddddd}
                   &CCSD(T)/     &CCSD(T)/     &CCSD(T)/     &  CCSD(T)/    &  CCSD(T)/    &  CCSD(T)/    & Microwave \\
                   &cc-pVTZ      &cc-pVTZ      &cc-pVTZ      &  cc-pVQZ     &  MTsmall     &  MTsmall     & Ref.\cite{Pea77} \\
$e^-$ correlated   & valence     & valence     & valence     & valence      & valence      & all          \\
                   &$r_e$        &$r_g-r_e$    &$r_z-r_e$    &  $r_e$       &  $r_e$       &  $r_e$       & $r_s$ \\
\hline                                                                                                    
& \multicolumn{7}{c}{CH$_2$NH}\\                                                                                              
\hline                                                                                                    
$r$(CN)            &   1.27746   &   0.00626   &   0.00576    &   1.27416    &   1.27331    &   1.27077   &    1.273 \\
$r$(NH)            &   1.02168   &   0.02040   &   0.00863    &   1.02000    &   1.02033    &   1.01912   &    1.021 \\
$r$(CH) cis        &   1.09236   &   0.02164   &   0.01088    &   1.09148    &   1.09168    &   1.09033   &    1.09 \\
$r$(CH) trans      &   1.08815   &   0.02141   &   0.01060    &   1.08728    &   1.08733    &   1.08602   &    1.09 \\
$\theta$(HNC)      & 109.510     &   0.131     &   0.271      & 109.934      & 109.640      & 109.784     &  110.4 \\
$\theta$(NCHcis)   & 124.624     &  -0.133     &   0.194      & 124.418      & 124.614      & 124.591     &  125.1 \\
$\theta$(NCHtrans) & 118.637     &  -0.320     &  -0.091      & 118.672      & 118.679      & 118.715     &  117.9 \\
\hline                                                                                                    
& \multicolumn{7}{c}{CH$_2$NH$_2^+$}\\                                                                                        
\hline                                                                                                    
$r$(CN)            &   1.27995   &             &              &   1.27693    &   1.27631    &   1.27393   &         \\
$r$(NH)            &   1.01657   &             &              &   1.01549    &   1.01550    &   1.01465   &         \\
$r$(CH)            &   1.08350   &             &              &   1.08271    &   1.08264    &   1.08136   &         \\
$\theta$(HNC)      &   121.510   &             &              &   121.501    &   121.525    &   121.534   &         \\
$\theta$(HCN)      &   119.404   &             &              &   119.394    &   119.425    &   119.457   &         \\
\end{tabular}

For an overview of the different types of molecular geometries and the mathematical 
relationships between them, see the review by Kuchitsu\cite{Kuc92}. $r_e$ is the
bottom-of-the-well equilibrium geometry, $r_z$ the position-averaged geometry in the
vibrational ground state, while $r_g$ is the geometry obtained in a gas-phase 
electron diffraction experiment.

\end{table}

\begin{table}
\caption{Computed and observed heats of formation of CH$_2$NH and CH$_2$NH$_2^+$,
and breakdown by components of the W2 computed heat of formation. All values in kcal/mol.\label{tab:heat}}
\squeezetable
\begin{tabular}{lcdddd}
                       & CH$_2$NH    &  C$_2$H$_4$   &trans-HNNH&  (a)   &  CH$_2$NH$_2^+$ \\
\hline
SCF/AV5Z               & 305.38   &  434.93 &  155.30  & -10.27 &  214.56 \\
SCF/AV$\infty$Z        & 305.44   &  434.98 &  155.31  & -10.30 &  214.64 \\
CCSD$-$SCF/AV5Z       & 122.30   &  117.90 &  128.52  &   0.91 &  115.93 \\
CCSD$-$SCF/AV$\infty$Z& 123.96   &  119.40 &  130.30  &   0.89 &  117.56 \\
(T)/AVQZ               &   8.23   &    7.18 &    9.66  &   0.19 &    6.61 \\
(T)/AV$\infty$Z        &   8.51   &    7.45 &    9.97  &   0.20 &    6.84 \\
inner-shell corr.      &   1.53   &    2.27 &    0.74  &  -0.03 &    1.70 \\
scalar relativistic    &  -0.34   &   -0.33 &   -0.31  &   0.02 &   -0.41 \\
spin-orbit coupling    &  -0.08   &   -0.17 &    0.00  &   0.00 &   -0.08 \\
TAE$_e$                & 439.02   &  563.64 &  296.04  &  -9.18 &  340.23 \\
ZPVE(scaled B3LYP)$^b$ &  24.59   &   31.48 &   17.50  &  -0.10 &   33.46 \\
anharmonic ZPVE        &  24.69$^c$&  31.52$^d$& 17.53$^e$ &  -0.17 &  --- \\
TAE$_0$                & 414.43   &  532.16 &  278.54  &  -9.08 &  306.77 \\
Previous benchmark     &          &  531.89\cite{c2h4tae} &   278.73\cite{n2h2} &\\
W2 \hofzero            &  22.98   &   14.33 &   49.78  &  -9.08 &  182.27 \\
W2 \HOF                &  21.08   &   12.28 &   48.07  &  -9.10 &  179.40 \\
W2h                    &  21.07   &   12.21 &   48.07  &  -9.07 &  179.41 \\
W1                     &  20.82   &   11.90 &   47.80  &  -9.03 &  178.96 \\
W1h                    &  20.86   &   12.12 &   47.80  &  -9.10 &  178.99 \\
Expt.                  &  26.4$\pm$3.2\cite{DeF78}, 25$\pm$3\cite{Gre88},&   12.52$\pm$0.12\cite{Gurvich} & $\geq$47.1$\pm$0.5$^g$ && 178$\pm$1\cite{Los81},\\
                       & 21$\pm$4\cite{russians,Hol92}, $\leq$22$\pm$3\cite{Hol92},&&&&179.7\cite{Ham99}\\
                       & 20.6$\pm$2.4$^f$\\
\end{tabular}

(a) reaction energy of  (1/2) C$_2$H$_4$+ (1/2) trans-N$_2$H$_2$$\rightarrow$ 
CH$_2$=NH with zero-point or temperature corrections following descriptions
in the first column

(b) B3LYP/cc-pVTZ harmonic frequencies scaled by 0.985, as prescribed in 
Ref.\cite{w1}.

(c) CCSD(T)/cc-pVTZ quartic force field, this work. 

(d) CCSD(T)/cc-pVQZ harmonics with CCSD(T)/cc-pVTZ anharmonicities, Ref.\cite{c2h4more}. 
At CCSD(T)/cc-pVTZ level 31.50 kcal/mol. Best estimate
in that reference is 31.59 kcal/mol.

(e) CCSD(T)/cc-pVQZ, Ref.\cite{n2h2}. At CCSD(T)/cc-pVTZ level 17.49 kcal/mol.

(f) from G2 reaction energies for 10 reactions, and expt. thermochemical
data for auxiliary species\cite{Smi92}

(g) From thermal correction in this work and $\Delta H^\circ_{f,0}\geq$48.8$\pm$0.5 kcal/mol
in H. Biehl and F.Stuhl, \JCP{100}{141}{1994}
\end{table}

\end{document}